\documentclass{xray_symp}
\input psfig.sty
\usepackage{graphics}
\def\R{{\it ROSAT}~}
\def\hi{H{\sc i}~}
\def\RAS{\R all-sky survey~}
\begin{document}
\title{Soft X-rays from High-Velocity Clouds}
\author{J. Kerp\inst{1}, P.M.W. Kalberla\inst{1}, M.J. Freyberg\inst{2},
Dap Hartmann\inst{1} \and W.B. Burton\inst{3}}
\institute{
Radioastronomisches Institut der Universit\"at Bonn
Auf dem H\"ugel 71, D-53121 Bonn, Germany
\and  
Max-Planck-Institut f\"ur extraterrestrische Physik, Postfach 1603, D-85740 Garching, Germany
\and
Sterrewacht Leiden, P.O. Box 9513, NL-2300 RA Leiden, The Netherlands}
\authorrunning{Kerp et al.}
\titlerunning{Soft X-ray from high-velocity clouds}
\maketitle

\begin{abstract}
The \R mission gave us the unique possibility to study the galactic interstellar
medium by ``shadowing'' observations on arcmin angular resolution level.
The high sensitivity of the \R PSPC in combination with the large field of
view allows the detailed study of individual clouds, while the \RAS is the
main source for studies of the large-scale intensity distribution
of the diffuse soft X-ray emission.

We show that {\em neutral} hydrogen clouds are associated with
soft X-ray emission, $E_{\rm X-ray}\,\leq\,\frac{3}{4}$\,keV.
These neutral hydrogen clouds, known as high-velocity clouds (HVCs),
are located partly within the galactic halo, and partly
in intergalactic space.
The \R detection of soft X-ray emission from some of these HVCs is the {\em first}
detection of HVCs in emission other than \hi 21-cm line radiation.

\end{abstract}

\section{Introduction}
The standard method for the analysis of an individual X-ray source is the subtraction
of the background from the observed total count rate.
This approach overcomes the main uncertainties introduced by the so-called non-cosmic X-ray
background radiation as well as residual contamination by scattered solar X-rays.
However, if the background radiation itself is the main interest, one needs excellent detectors
and a clear understanding of the residual contamination of the data, which is not 
automatically rejected by the detector electronics or the raw-data processing.

The \R PSPC was a superb detector to investigate the background emission, especially
for the analysis of the soft X-ray emission within the $\frac{1}{4}$\,keV band.
Towards the high galactic latitude sky,
the product of the neutral hydrogen column density and the interstellar
photoelectric absorption cross section reaches at least unity within the
 $\frac{1}{4}$\,keV energy band.
Accordingly, $\frac{1}{4}$\,keV soft X-ray shadows of individual clouds are expected for clouds
located in front of an extended X-ray source.

The first maps of the diffuse soft X-ray background (McCammon et al. \cite{mccbur}) did
not resolve individual cloud shadows, but confirmed the general negative correlation
between the soft X-ray background intensity distribution and the galactic neutral hydrogen
distribution (Bowyer, Field and Mack \cite{bowfie}). Moreover, the Wisconsin survey established the
existence of a $T\,\simeq\,10^6$\,K plasma in the local void of neutral matter,
the so-called local hot bubble (LHB, Cox \& Reynolds \cite{coxrey}).

One of the first major discoveries of the \R mission was the detection of the ``Draco shadow''
by Snowden et al. (\cite{snomeb}). The $\frac{1}{4}$\,keV shadow of this cloud, which is located
outside the LHB, gave evidence for an X-ray emitting plasma beyond the LHB.
This X-ray emitting plasma within the galactic halo was attributed
to an individual hot spot (Burrows \& Mendenhall \cite{burmen}).
During the course of the \R mission, more $\frac{1}{4}$\,keV X-ray shadows of neutral clouds were
discovered (i.e. Kerp et al. \cite{kerher}, Snowden et al. \cite{snomcc}, Herbstmeier et al. \cite{herker}).

There has been a lot of discussion about whether the distant X-ray emission is associated with a smoothly
distributed X-ray emitting plasma within the galactic halo (Kerp et al. \cite{kerbur}, Pietz et al. \cite{pieker},
Freyberg \cite{freyber}) or whether it is the superposition of individual scattered hot spots
(Snowden et al. \cite {snoegg}, Shelton \cite{shelt98}).

In this paper we will demonstrate that a smooth, constant
intensity distribution of the halo X-ray plasma, quantitatively fits the
observed soft X-ray background (SXRB) intensity distribution (Sect.\,2)
very well.
This modelling of the SXRB intensity distribution allows us to disclose
sky areas of too faint or too bright soft X-ray emission. This approach provides evidence that some
HVCs are associated with $\frac{1}{4}$\,keV soft X-ray emission (Sect.\,3).

\section{Modelling the SXRB}
Figure~\ref{radtrans} shows the application of our model to the SXRB intensity distribution, step-by-step.
The intensity profiles (Fig.~\ref{radtrans} {\it left\/}) are extracted from the corresponding maps (Fig.~\ref{radtrans} {\it right\/}).
The arrows mark the (constant) galactic latitude position of the slices, while the intensity profiles are a function of
the galactic longitude.
Fig.~\ref{radtrans}{\bf a)} shows the assumed constant intensity distribution of the X-ray halo. The unabsorbed 
intensity level seen on the y-axis of the intensity profile is $I_{\rm halo}\,=\,25\cdot 10^{-4}\,
{\rm cts\,s^{-1}\,arcmin^{-2}}$.

In panel {\bf b)}, the soft X-ray attenuating column density distribution of $N_{\rm HI}$ is shown.
Black indicates low column densities, and white high column densities.
On the left the corresponding slice at $b\,=\,40\degr$ shows that the column density, $N_{\rm HI}$, varies between
$1\,{\rm to}\,5\cdot 10^{20}\,{\rm cm^{-2}}$.

Panel {\bf c)} shows the modelled $\frac{1}{4}$\,keV intensity distribution of $I_{\rm halo}$ (panel {\bf a)}) absorbed
by the galactic interstellar matter ($N_{\rm HI}$; panel {\bf b)}). 
The longitude profile on the left $I_{\rm halo}$ is attenuated by a factor of 3 to 5. The minimum X-ray
intensity level of $I_{\rm X-ray}\,=\,5\cdot 10^{-4}\,{\rm cts\,s^{-1}\,arcmin^{-2}}$ is a superposition of
the attenuated $I_{\rm halo}$ and the foreground $I_{\rm LHB}\,=\,3\cdot 10^{-4}\,{\rm cts\,s^{-1}\,arcmin^{-2}}$.
Thus, towards high $N_{\rm HI}$ regions, only 40\% of the detected X-ray photons have their origin
in the galactic halo plasma; the rest originates from the LHB. The
unabsorbed intensities are $I_{\rm halo}\,/\,I_{\rm LHB}\,\geq\,8$.

Finally, in panel {\bf d)}, we present the \RAS data for this $25\degr\,\times\,25\degr$ region.
The left panel shows a superposition of the observed intensity-profile at $b\,=\,40\degr$ from the \RAS data and the modelled intensity profile (thick line).
Considering the uncertainties in the \RAS data of about $\Delta I_{\rm X-ray}\,\simeq\,0.8\cdot 10^{-4}\,
{\rm cts\,s^{-1}\,arcmin^{-2}}$, the observed and modelled intensity profiles 
agree very well.

Using this method, Kerp et al. (\cite{kerbur}) analysed four large areas of the X-ray sky to search for signatures of HVCs.
They proved statistically that the modelled and observed map intensity distributions match perfectly.

\section{X-rays from HVCs}

Kerp et al. (\cite{kerbur}) analysed four large fields using the method
described above, to search for signatures of HVCs in the \RAS data.
In Fig.\,\ref{hvcivc}, we show a mosaic of the \RAS data towards the 
HVC complex C.
The \RAS field described in Fig.\,\ref{radtrans} is the lower right area of the mosaic.
Because of the well modelled intensity distribution of the ``normal'' SXRB (Fig.\,\ref{radtrans}), we can now examine areas of
excess soft X-ray emission by subtracting the model from the observed SXRB 
intensity distribution.
Excess SXRB emission is defined here more significant as residuals $>5\sigma$ (Fig.~\ref{hvcivc}).
Most of the SXRB emission is modelled well, and no large-scale excess emission
is found.
The distribution of the soft X-ray excess is patchy, with the most prominent one located
close to the position of the Draco cloud ($l\,\sim\,90\degr,~b\,\sim\,39\degr$).
Overall, the galactic X-ray halo has a {\em smooth} large-scale-intensity
distribution, while close to some of the HVCs bright excess soft X-ray patches are located. 

Superimposed on the excess soft X-ray emission (grey-scale in Fig.~\ref{hvcivc}) as contours, is the \hi column
density distribution of HVC complex C.
Although there is no one-to-one positional correlation between the
excess X-ray emitting areas and the HVCs,
the places where SXRB excesses are seen do coincide with the locations of HVCs.

Kerp et al. (\cite{kerbur}) found a total of 12 excess X-ray emitting areas. Integrated across the whole
excess area, the detected excess $\frac{1}{4}$\,keV energy is about $E_{\rm det}\,=\,1\cdot 10^{-10}{\rm erg\,s^{-1}
cm^{-2}}$.
Accordingly, the unabsorbed $\frac{1}{4}$\,keV X-ray flux is about $E_{\rm rad}\,=\,5\cdot 10^{34}{\rm erg\,s^{-1}}$ ($E_{\rm rad}\,\propto\,\frac{D^2}{\rm [kpc]}$).
The maximum in the \R $\frac{1}{4}$\,keV data is $E_{\rm rad}\,\sim\,10^{36}{\rm erg\,s^{-1}}$, which is only a tiny
fraction of the available kinetic energy of HVC complex C ($E_{\rm kin}\,\sim\,10^{53}{\rm erg}$).

The emission measure is about $EM\,=\,0.03\,{\rm cm^{-6}\,pc}$, which is roughly 6 times higher than
the emission measure of the galactic halo plasma ($EM_{\rm halo}(b\,=\,90\degr)\,=\,5\,10^{-3},{\rm cm^{-6}\,pc}$, Pietz
et al. \cite{pieker}).
The high excess emission measure implies, that the HVC material itself may be partly ionised and heated
up to temperatures of about $T\,=\,10^{6.2}$\,K (Kerp et al. \cite{kerpII}).
If we compare the volume densities of the electrons and of the neutral
hydrogen atoms, we find that $n_{\rm e}/n_{\rm HI}\,\simeq\,0.3\,\frac{\sqrt{D}}
{\rm [kpc]}$.
If we assume that HVCs are ``pure'' hydrogen clouds, this
ratio is an estimate of the ionisation fraction of the atomic hydrogen ($N_{\rm HII}/N_{\rm HI}$). 
We can invert the electron-to-neutral-hydrogen atom density ratio to calculate the
distance for a completely ionised hydrogen HVC to be $D\,>$\,11\,kpc.
This distance limit is close to the observationally determined upper distance limit 
(van Woerden et al. \cite{woewak}) and those based on theoretical considerations (Blitz et al. \cite{blitz98}).
We can interpret this high ionisation of HVCs in two ways:
First, the determination of the emission measure may be in error, because the collision equilibrium
plasma models do not fit the real situation.
Second, a significant fraction of the HVCs hydrogen is {\em not} neutral, so the HVCs may have a larger extent and contain more 
mass than the neutral hydrogen indicates. 
This might explain the obvious positional
offset between the bright soft X-ray enhancements and the neutral hydrogen column
density distribution.
Also, some of the ${\rm H\alpha}$ emission maxima are found to be not positionally coincident with
$N_{\rm HI}$ maxima (i.e. MI-cloud position 6a, Tufte, Reynolds and Haffner \cite{tufrey}).

Further studies of the X-ray emission of HVCs will be carried out by correlating of 
the \RAS and the Leiden/Dwingeloo survey, as well as through future XMM observations.
\begin{figure*}
\caption{
The soft X-ray radiation transport of $\frac{1}{4}$\,keV photons
through the galactic interstellar medium.
{\it \bf Right:\/} {\bf a)} The assumed {\em constant} intensity distribution
of $I_{\rm halo}$. {\bf b)} The distribution of the $\frac{1}{4}$\,keV 
absorbing interstellar matter, traced by the \hi 21-cm column density 
distribution (Hartmann \& Burton \cite{harbur}).
{\bf c)} The SXRB intensity distribution of the constant $I_{\rm halo}$
intensity (panel a)) after the transition through the absorbing medium 
(panel b)).
{\bf d)} The \RAS data of the same region. 
The comparison of the maps in panels c) and d) shows an excellent agreement between
the model and the observations.
{\it \bf Left:\/} Intensity profiles at $b\,=\,40\degr$ from the maps on the 
right.
Superimposed on the intensity profile of the \RAS map in panel {\bf d)}, is the
intensity profile from the model in panel {\bf c)}.
The maps on the right cover an area of about $25\degr\,\times\,25\degr$.
The values of $I_{\rm X-ray}$ and $N_{\rm HI}$ are given by the
labelling of the left-hand side panels. The $I_{\rm LHB}\,=\,(2.8\,\pm\,0.5)\,
10^{-4}\,{\rm cts\,s^{-1}\,arcmin^{-2}}$, which is nearly an order of magnitude
fainter that $I_{\rm halo}$ (panel {\bf a)}).
}
\label{radtrans}
\end{figure*}

\begin{figure*}
 \caption{
Mosaic showing the $\frac{1}{4}$\,keV and \hi data towards HVC complex C.
{\bf Top left:} The \R $\frac{1}{4}$\,keV map towards HVC complex C.
{\bf Top right:} The \hi column density map of the same field of view, extracted
from the Leiden/Dwingeloo survey within the velocity range $v_{\rm LSR}\,\in\,[-100;+100]\,{\rm km\,s^{-1}}$. 
{\bf Bottom:} The positional correlation of excess 
 $\frac{1}{4}$\,keV emission and the HVC ({\bf left}) and IVC ({\bf right}) $N_{\rm HI}$ distributions 
 towards the HVC complex C. The images present the areas of excess soft 
X-ray emission in the significance range 4$\sigma$ (black) to 10$\sigma$ 
(white).  {\bf Bottom left:} The HVC $N_{\rm HI}$ distribution 
 ($ -450 < v_{\rm LSR} < -100\, {\rm km\,s^{-1}}$) superposed as contours 
 with $1\cdot 10^{19}\,{\rm cm^{-2}}\,\leq\, N_{\rm HI}\,\leq\,1\cdot 
 10^{20}\,{\rm cm^{-2}}$ in steps of $\Delta N_{\rm HI}\,=\,1\cdot 
 10^{19}\,{\rm cm^{-2}}$.
 {\bf Bottom right:} The IVC $N_{\rm HI}$ distribution ($-75 < v_{\rm LSR} < -25\, 
 {\rm km\,s^{-1}}$) superposed as contours with $5\cdot 10^{19}\,{\rm 
 cm^{-2}}\,\leq\, N_{\rm HI}\,\leq\,2\cdot 10^{20}\,{\rm cm^{-2}}$ in steps 
 of $\Delta N_{\rm HI}\,=\,2.5\cdot 10^{19}\,{\rm cm^{-2}}$.
 The HVC $N_{\rm HI}$ distribution follows the orientation of the soft X-ray 
 enhancements. The $N_{\rm HI}$ maxima of the IVCs are not positionally coincident with 
 excess X-ray areas, except near $l\,\sim\,102\degr$, $b\,\sim\,37
 \degr$, close to the Draco nebula, and near $l\,\sim\,118\degr$, $b\,
\sim\,42\degr$. Both IVCs are located close to HVCs and perhaps link
both cloud populations. 
}
\label{hvcivc}
\end{figure*}


\begin{thebibliography}{}
\bibitem[1998]{blitz98}
 Blitz L., Spergel D.N., Teuben P.J., Hartmann Dap, Burton W.B., 1998, ApJ in press (astro-ph/9803251)
\bibitem[1968]{bowfie}
Bowyer C.S., Field G.B., Mack J.E., 1968, Nat 217, 32
\bibitem[1993]{burmen}
 Burrow D.W., Mendenhall J.A., 1993, Nat 351, 629

\bibitem[1987]{coxrey}
Cox D.P., Reynolds R.J., 1987, ARA\&A 25, 303

\bibitem[1997]{freyber}
Freyberg M.J., 1997, ``The physics of galactic halos'', Eds. Lesch H. et al. p. 117 

\bibitem[1997]{harbur}
Hartmann Dap, Burton W.B., 1997 ``Atlas of Galactic Neutral Hydrogen'', Cambridge University Press

\bibitem[1994]{herker}
Herbstmeier U., Kerp J., Moritz P., 1994, ``Reviews of Modern Astronomy'' No. 7, p. 151 

\bibitem[1993]{kerher}
Kerp J., Herbstmeier U., Mebold U., 1993, A\&A 268, L21

\bibitem[1999]{kerbur}
Kerp J., Burton W.B., Egger R., Freyberg M.J., Hartmann Dap, Kalberla P.M.W., Mebold U., Pietz J., 1999, A\&A in press (astro-ph/9810307)

\bibitem[1996]{kermac}
Kerp J., Mack K.H., Egger R., Pietz J., Zimmer F., Mebold U., Burton W.B., Hartmann Dap, 1996, A\&A 312, 67

\bibitem[1998]{kerpII}
Kerp J., Pietz J., Kalberla P.M.W., Burton W.B., Egger R., Freyberg M.J., Hartmann Dap, Mebold U., 1998, IAU Coll. 166 ``The local bubble and beyond'', Eds. Breitschwerdt D., Freyberg M.J., Tr\"umper J., Lecture Notes in Physics 506, 457

\bibitem[1983]{mccbur}
 McCammon D., Burrow D.W., Sanders W.T., Kraushaar W.L., 1993, ApJ 269, 107

\bibitem[1996]{pietz96}
Pietz J., Kerp J., Kalberla P.M.W., Mebold U., Burton W.B., Hartmann Dap, 1996, A\&A 308, L37

\bibitem[1998]{pieker}
Pietz J., Kerp J., Kalberla P.M.W., Mebold U., Burton W.B., Hartmann Dap, 1998, A\&A 332, 55

\bibitem[1998]{shelt98}
Shelton R.L., 1998, IAU Coll. 166 ``The local bubble and beyond'', Eds. Breitschwerdt D., Freyberg M.J., Tr\"umper J., Lecture Notes in Physics 506, 443

\bibitem[1998]{snoegg}
Snowden S.L., Egger R., Finkbeiner D.P., Freyberg M.J., Plucinsky P.P., 1998, ApJ 493, 715

\bibitem[1993]{snomcc}
Snowden S.L., McCammon D., Verter F., 1993, ApJ 409, L21

\bibitem[1991]{snomeb}
 Snowden S.L., Mebold U., Hirth W., Herbstmeier U., Schmitt J.H.M.M., 1991, Sci 252, 1529

\bibitem[1998]{tufrey}
Tufte S.L., Reynolds R.J., Haffner L.M., 1998, ApJ 504, 773

%\bibitem[1998]{wang98}
%Wang Q.D., 1998, IAU Coll. 166 ``The local bubble and beyond'', Eds. Breitschwerdt et al., Lecture Notes in Physics 506, 503
%
%\bibitem[1997]{wakwoe}
%Wakker B.P.,van Woerden H., 1997, ARA\&A 35, 217

\bibitem[1998]{woewak}
van Woerden H., Wakker B.P., Schwarz U.J., Peletier R., Kalberla P.M.W., 1998, IAU Coll. 166 ``The local bubble and beyond'', Eds. Breitschwerdt D., Freyberg M.J., Tr\"umper J., Lecture Notes in Physics 506, 467

\end{thebibliography}
\end{document}